**Age, Sex, and Genetic Architecture of Human Gene Expression in EBV Transformed Cell Lines**

Manuel A. Rivas[1,2,4], Mark J. Daly[2,3], Itsik Pe'er[4]

1 Department of Mathematics, Massachusetts Institute of Technology, Cambridge, Massachusetts 02139, USA

2 Program in Medical and Population Genetics, Broad Institute of Harvard and Massachusetts Institute of Technology, Cambridge, Massachusetts 02142, USA

3 Center for Human Genetics Research and Department of Molecular Biology, Massachusetts General Hospital, Boston, Massachusetts 02114, USA

4 Department of Computer Science, Columbia University, New York, New York 10027, USA

Corresponding authors:

| **Mark J. Daly** | **Itsik Pe'er** |
|---|---|
| Massachusetts General Hospital | Columbia University |
| Center for Human Genetics Research | 505 Computer Science Building |
| 185 Cambridge St | 1214 Amsterdam Avenue, Mailcode: 0401 |
| Boston MA 02114 | New York NY 10027-7003 |
| USA | USA |
| Tel: 617-643-3290 | Tel: 212-939-7135 |
| mjdaly@chgr.mgh.harvard.edu | ip2169@columbia.edu |

**e-mail addresses**

Manuel A. Rivas rivas@broad.mit.edu

Mark J. Daly mjdaly@chgr.mgh.harvard.edu

Itsik Pe'er ip2169@columbia.edu

Individual expression profiles from EBV transformed cell lines are an emerging resource for genomic investigation. In this study we characterize the effects of age, sex, and genetic variation on gene expression by surveying public datasets of such profiles. We establish that the expression space of cell lines maintains genetic as well as non-germline information, in an individual-specific and cross-tissue manner. Age of donor is associated with the expression of 949 genes in the derived cell line. Age-associated genes include over-representation of immune-related genes, specifically MHC Class I genes, a phenomenon that replicates across tissues and organisms. Sex associated genes in these cell lines include likely candidates, such as genes that escape X-inactivation, testis specific expressed genes, androgen and estrogen specific genes, but also gene families previously unknown to be sex associated such as common microRNA targets (MIR-490, V_ARP1_01, MIR-489). Finally, we report 494 transcripts whose expression levels are associated with a genetic variant in cis, overlapping and validating previous reports. Incorporating age in analysis of association facilitates additional discovery of trans-acting regulatory genetic variants. Our findings promote expression profiling of transformed cell lines as a vehicle for understanding cellular systems beyond the specific lines.

**Author Summary**

It is well known that measurements of gene expression levels are useful in understanding the roles, pathways, and function of genes across the organism under study. Here we take advantage of recent datasets and methods that allowed us to address several questions about aging, sex, and genetics, such as: Are age differences reflected in human gene expression levels in more than one tissue, is information relating to the individual reflected in EBV transformed cells, and are there genetic variants that influence expression of genes? We demonstrate that there are immune specific genes that are associated with chronological age in both EBV transformed cell lines and kidney, that the common sex associated genes are also reflected in these cell lines and identify new gene groups that were not suspected to be sex associated genes, and find variants across the genome that are indeed influencing levels of gene expression in this tissue. Future implication includes studying aging, sex, and genetic influences in expression levels of other human cell lines and tissues.

Genomewide association studies and expression profiling of disease case and control individuals are powerful strategies to properly evaluate the genes that impact human health. Recent studies have attempted to demonstrate a general overlap between SNPs associated to disease traits and those associated to RNA levels in accessible tissues with limited success and new models were created to approach a problem that integrates genetics, gene expression, and disease [1,2]. Genetic genomics holds the promise to link these paradigms by dissecting all the heritable and environmental factors that regulate expression-mediated processes in the cell [3]. Indeed, coupled whole genotype and expression data for human lymphoblastoid cell lines have recently been analyzed by classical methods to map numerous QTLs by linkage in pedigrees [4,5,6] and association in unrelateds [7,8,9]. Integration of genetic and gene expression information in the extended pedigree setting has shown association of only a limited number of transcript levels to variation in the genome [10].

The dependencies of a disease trait on age, sex, and heritability from each parent are routinely evaluated for consideration in design and analysis of genetic studies of such a trait. The reported association findings motivate exploration of the connections between these factors and gene expression phenotypes. A major challenge has to do with the human system for these studies. The transcriptome-wide effects of disease on expression profiles [1] and the infiltration of multiple cell types [2] highlight the issues confronted by genetic- genomic investigation in primary tissue. We focus on genetic genomics in lymphoblastoid cell lines that are Epstein-Barr virus transformed. While these lines maintain the DNA of the contributing individual, immortalization, tissue specificity, lab culture conditions, cellular phenotypes such as growth rate and metabolic state [11], as well as RNA levels of the Epstein-Barr virus used to transform the cells may obliterate individual signatures of gene expression, and such effects have been largely ignored in the association analysis of cell-line expression levels as quantitative genetic traits for genetic association.

We thus sought to assess the effects of age, sex, and heritability on transcript levels themselves, as reflected by EBV-transformed cells. We hypothesized that age and sex would be covariates of expression for many genes, motivating refined analysis of genetic association for these transcripts. We consider public data from CEPH extended pedigrees that include expression profiles as well as microsatellite and SNP genotypes [12,13]. Such data provide ample transmissions to assess heritability [4] and span a wide range of ages [Fig 1a]. In terms of genetics, the highly dense, Phase II HapMap [14,15] allows for a finer scale interpretation of the linkage signals previously reported [4] in pedigrees that overlap with the HapMap samples. Here we have used transcriptional profiles of five tri-generational pedigrees and five trios and genotypes of 2.2 million SNPs genotyped by the HapMap Consortium. Furthermore, we use transcriptional profile of 270 individuals in the four HapMap populations to assess sex specific signatures in gene expression [7,8]. Such mixed data, of a microsatellite-typed pedigree, array-typed unrelateds and expression profiles have recently shown to expose large scale transcription signatures of disease and genetics [1], and may show much more using methodologies presented here to leverage all samples.

**Results**

**Age architecture of human gene expression**

Few studies have tried to systematically evaluate the relationship of chronological age and sex with expression phenotypes across multiple tissues [16,17,18]. Recent studies in mice have catalogued changes in gene expression that are associated with these covariates and demonstrated patterns that were conserved across tissues and other patterns that were specific [17,18]. However, no such comparison has been made across tissues in a systematic fashion with modest sample sizes and with a broad range of chronological age in a human system. The insight into cross tissue conservation of patterns demonstrates the importance of systematically addressing this problem in future work.

We model the age architecture of human gene expression by taking advantage of the broad age distribution available in the extended-pedigree setting with three generations and 142 samples available to determine the relationship between chronological age and lymphocyte expression levels [Fig. 1a,1b][Eq. 1]. 2991 genes are significantly (p=0.01) affected by donor age (FDR=0.08). Taking a genomewide approach, we examined the functional annotation of age-associated transcripts and found enrichment for immune-related genes. A similar phenomenon has been observed in systems other than human lymphoblastoid cell lines [16,19,20,21,22,23,24], alleviating many technical concerns of a statistical or experimental artifact. Seeking to further validate age-associated expression on a gene-specific resolution, we compared our analysis to analogous analysis of human kidney data [16]. These studies examine different tissues and their small sample sizes ($N_{kidney}$=74,$N_{lymphoblastoid}$=142) limit their power to detect association with age, therefore complete overlap of associated genes is not expected. Yet, we find significant overlap between the sets of genes whose expression increases in both datasets (57 genes, $\chi^2$ = 37.52, p-value = 9.04e-10, see Figure 1c)[Supp Table 2]. Functional analysis reveals genes of the MHC-class I to comprise 10% of this overlapping set including HLA-A, B, E, F, and HUMMHCW1A Cw1 antigen (which maps to HLA-C gene in the genome) and many MHC processing and antigen presenting genes. As previously described in similar studies of aging in mice, different cell types, specifically murine hippocampus, motorneurons, and T cells maintain the signature of an increased expression level of MHC Class I genes in older mice [19,20,21,22,23,24]. We note that a major concern regarding immune-genes in the kidney had to do with potential tissue contamination with immune system cells [16]. It was therefore suggested that an increased number of immune cells were present in the kidney of older individuals, resulting in an age-related increase in abundance of all genes that are expressed specifically in those cells, without a true change in expression level in any particular cell type. A related potential source of artifactual age-associated expression of immune-related genes could be due to increase overall expression within immune cells contaminating the kidney tissue. The data reported here from lymphoblastoid cell lines proves the overexpression of this set of genes is genuine. This signature of aging and the enrichment of transcripts associated with chronological comes as a surprise since the cell lines have been immortalized. The conserved signature across kidney and EBV transformed cell lines demonstrates the utility of cross tissue comparison of gene expression signatures. No significant overlap is detected across these tissues beyond this set of genes, nor among genes that decrease in expression with age (8 genes, $\chi^2 = 0.973$, p-value = 0.3239). We predict that among the factors that can attribute to this absence of overlap include small sample sizes used in each individual study and that age attributes only a modest effect to expression change in either tissue being studied.

We considered hierarchical clustering of all 951 transcripts associated with old-age expression. [Figure 1d] shows the tightest ($r^2 = .78$) cluster in this data. It includes 11 genes including Apolipoprotein E (APOE), which validates in both human kidney [16] and murine hippocampus data in mice [22] as age associated. As common genetic variation in APOE is associated to numerous late-onset conditions (e.g. cardiovascular disease [25], dislipidemia [26], colorectal cancer [27], Alzheimer's disease [28] as well as longevity [29]) we sought to further understand the expression signature of this cluster through genetic lens. We scanned the data for potential genetic mediators of the age-related expression differences observed. We used dense SNP genotype data available by the HapMap for individuals from two generations of CEPH. To maximize statistical power we increase sample size by inferring SNP genotypes for their third-generation descendants based on available microsatellite and SNP data [10] (see Methods). Given the derived sample dataset with overlapping genotype and gene expression we fit a model that would test for age by genotype interaction in the expression dataset with the 69 samples available (see Methods). We report 11 variants across 4 loci, as showing significant (p-value < 1e-6) age by genotype interaction in the dataset. In particular, these variants include rs10928136 a SNP located in the intronic sequence of LRP1B, a receptor of APOE [Figure 1d,2, Supp Table 3] [30]. Further investigation and replication is warranted as it could potentially reveal some novel insights into the regulation of APOE.

**Sex architecture of human gene expression**

Sex is perhaps the most obvious biological division of humans, with manifestations ranging from genetics to appearance to health [31,32]. Understanding the partition of genome expression differences is important to understand the role of sex in genome science. While EBV transformed cell lines carry the sex specific karyotype, manifestation of such expression differences in these cells is yet to be evaluated. To properly quantify these differences we use existing HapMap cell line measurements in the GeneVar dataset [7] and derive common signatures that are shared across all four populations. Furthermore, we investigate the enrichment of genes with sex derived expression using Gene Set Enrichment Analysis (GSEA).

To analyze changes in gene expression with sex in human lymphocytes, we used 90, 90, 45, 45 individual lymphoblastoid cell lines samples of CEPH, YRI, JPT, and CHB ancestry. To guard against statistical concerns of population stratification, we first took a conservative approach and analyzed each population separately. We used a linear regression model to identify genes that showed a statistically significant change in expression with sex (Methods). Despite the limited sample size, for 3 (YRI, JPT, CHB) of the 4 populations we saw a significant (p-value < 1e-10) enrichment of transcripts whose expression is associated with sex, compared to random expectation [Figure 3]. We note that CEU/CEPH cell lines were collected much earlier than the other cell lines from the pertaining individuals and might be an explanation for the lack of overdispersion of sex associated genes in that population. This observation would spark concern about CEPH cell lines. However, for both males and females, there is significant overlap between the genes that are expressed in that specific sex across the board for the four populations (including CEPH) in all chromosomes [Supp Table 1]. Hence, this only indicates that our power is limited for the CEPH cell lines to robustly detect any gender specific expression changes and combining all four populations would leverage detection. As expected, transcripts associated with males in one population do not significantly overlap those with opposite association in other direction, suggesting the overlap in the same direction is indeed significant, rather than an artifact of the scoring method. When we restrict the analysis to the autosomes we observe significant overlap in 5 of the 6 pairwise comparisons, implying a signature of autosomal genes that are expressed differentially between males and females.

Encouraged by our ability to detect significant sex association even with only dozens of samples in each analysis of a single population, we sought to enhance power by combining the statistic of the four populations. We merged the 3,637 transcripts [expected = 2956] which were preferentially expressed in males in all four populations and merged their p-values using Fisher's exact test method (see methods). The list of genes ranked by combined p-values was examined for enriched gene sets [33]. Similar analysis was applied to the set of 2,567 female-overexpressed.

**Gene Set Enrichment Analysis**

In the context of obvious biological sex differences in gene expression (sex chromosomes) we would not expect to observe X chromosome genes to be higher expressed in males or Y chromosome genes to be expressed at all in females. Single gene analysis at a = .01 results in a single gene list that of 3 X-chromosome genes that overlap in two or more populations and are higher expressed in males than in females (NLGN4 [autism related], SCML1, PNMA3) , and a single gene on the Y chromosome, PRKY. PRKY is located on chromosome Y, near the boundary of the pseudo autosomal region. It is known that abnormal recombination between this gene and a related gene on chromosome X is a frequent cause of XX males and XY females [32]. Due to the high sequence similarity with its homolog, PRKX, this observation may be attributed to probe similarity.

**Cytogenetic gene sets** on sex chromosomes whose expression is found to be sex-dependent include ChrYq11 for males and ChrXp11 and ChrXp22 for females. The overexpression of autosomal Chr18q22 genes is also found to be enriched in males [Supplementary Table Website]. **Curated gene sets** that are observed to be overexpressed in males include testis expressed genes, genes upregulated in pulpal tissues from extracted caries, and ribosomal protein genes. In females, we observe genes escaping X inactivation to be upregulated, genes involved in the androgen and estrogen metabolism pathway. **Motif gene sets** that are significantly upregulated in males (7 sets, nominal p-value < .01) and females (2 sets, nominal p-value<0.05) include microRNA motifs; *mir-489*, *mir-342*, *mir-500*(m), and *mir-490* (f). Lastly, 14 out of 30 **computational, cancer-associated gene sets** are significantly overexpressed in males at nominal p-value < .01.

**Genetics of gene expression**

Continuing our analysis of the genetic and non-genetic factors that influence gene expression levels, we surveyed the relationship between SNP variation and gene expression levels. We use CEPH family extended pedigree expression data previously generated along with sparse microsatellite data, and highly dense imputed and typed HapMap Phase II SNP data to fine map expression traits. An obvious concern is that the quantitative traits under study of the individuals in the setting of this study are not truly independent and it would not be suitable to treat them as such. However, genomic control [38,39,40] [(See Methods) permits dependent outcomes to be treated as independent observations and the computed statistics can be adjusted, taking care of any relatedness and/or inherent bias in the data.

Of obvious interest is to assess the power gains or loss in analysis of genetics of gene expression by adjusting for the covariates (age and sex specifically) that we examine for relationships with expression traits. In our study we scan all expression phenotypes and their neighboring SNPs in a 5Mb window for cis-acting variants and apply genomic control by selecting 100 random SNPs with MAF > 0.15 drawn across the entire genome.

We focus on a linkage peak on chromosome 6, reported to affect the expression of six transcripts [4] that reside on different chromosomes. We used HapMap-typed and inferred SNP genotypes to fine map this linkage peak. and find rs2157337 to be associated with the expression of these six genes (Supp Table 4). Evaluation of this effect in HapMap expression profiles replicated HLA-DRB3 association with this SNP, which is 1.4MB away from the transcript (Supp Table 4).

Analysis of cis regulatory variation allows for a systematic evaluation of power to replicate associations across this and the GENEVAR study as such analysis enjoys a lighter burden of multiple testing and therefore numerous such associations have been reported [6,7,8,9]. Of the 23,880 transcripts that were tested, we detect significant (p-value < 1e-5) association in CEPH pedigrees with at least one SNP in cis for 494 transcripts after genomic control without adjusting for age and sex, and 477 after adjusting for age and sex [Supp Table 5a,b & Supp Fig. 1], 339 of which overlap for both models. Each of the gene sets that did not overlap between these two models (155 and 138 genes, respectively) had a subset of 22% of its members (35/31 genes, respectively) map well to the GeneVar dataset, a third of which (12/11 genes, respectively) were validated by the GeneVar analysis. This validation rate can be contrasted with the 339 snp-gene pairs that agreed with both models, where 83 mapped well and 67 snp-gene pairs were validated. The results suggest similar power for both models of association, with and without correcting for age/sex, with different genes better detectable by different analyses[1].

Analysis of regulation by cis-variation also allowed examination of the results in the context of disease association, as pointed out by recent studies [35,36] In particular, we also find rs6969930 to be a cis acting determinant of IRF5 gene expression phenotype (p-value = 1.11e-16, see methods) that confers risk to systemic lupus [35], and rs7216389 a cis acting determinant of LOC51242 or ORMDL3 gene expression (p-value = 3.68e-06) that confers risk to childhood asthma [36].

**Discussion**

The regulation program of a cell ranges from constant expression of certain transcripts to variable transcript levels, responsive to stimuli and conditions. Individual factors affecting transcription profiles are unique in affecting steady-state measurements, yet facilitate examination of inter-sample variation. In this work we surveyed factors affecting individual expression of a large fraction of the transcriptome in a tissue available for diagnostics, reporting pervasive effects of donor age, sex, and genetic variation on expression levels in lymphoblastoid cell lines. While genetics provides an individualized mechanism of information transmission to explain cis-acting variation and sex-dependent expression, and the latter is supported by chromosomal location of the detected transcripts, the involvement of miRNA families in sex-specific expression is suggestive of a mechanism for this regulation.

It is less obvious how age affects many transcripts in such cell lines under controlled conditions, may be consistent with epigenetic mechanisms. Specifically, age-related MHC-I expression across tissues alleviates the concern of artifacts, and validations across model systems rules out trends specific to human aging, like history or changes in health and immunization policy.

Together, our observations are points in favor the use of cell line expression and genetics as a model system for regulation in the entire organism, and motivate further investigation of similar datasets. Furthermore, bearing in mind that both expression and SNP arrays revolutionized disease research by providing genomewide observations on potential effects and causes of clinical conditions, the combined analysis of expression, genetics and disease is an attractive goal. Specifically, our results suggest that improved understanding of the individual-specific variables affecting transcriptional programs will facilitate better using expression profiles as intermediates in mapping connections between genetic variation and disease.

**Methods**

**Overall strategy of accumulating data and power for genetics of gene expression**

We seek to integrate genetics and other factors that might give insight into the sources of variation in gene expression measurements in human cell lines. Gene expression measurements of 23,880 transcripts were previously collected from lymphocytes in individuals from CEPH extended pedigrees [4]. We use HapMap Phase II genotypes from individuals in CEPH Trios 1345, 1346, 1358, 1375 and a Duo 1349. We integrate HapMap Phase II genotypes from complementary trios in CEPH families 1334,1340,1350,1362, and 1408 along with microsatellite and SNP data [12,15]. We perform imputation of 3M markers across all available sibships in the overlapping trios in the HapMap by integrating microsatellite and SNP data with our own implementation of the algorithm described in [10] and use 43 of those individuals for the current study. By doing so, we increase our sample size from 26 individuals to 69 individuals with densely typed genotypes and gene expression traits. Family sample sizes: 1334 (9), 1340 (8), 1350 (10), 1408 (12), 1362 (17). Age data for individuals in the sample is available for 65 individuals.

**Age association to gene expression levels**

To understand the effect of chronological age as a source of variation in gene expression levels across the entire set of transcripts we fit [Eq. 1] to 142 individuals in the gene expression array with pertinent age data available

(1) $$Y_{ij} = \beta_{0j} + \beta_{1j} Age_j + \beta_{2j} Sex_j + \sum_{k=3}^{15} \beta_{kj} Family_j + \varepsilon$$

where the family component is added to adjust for any inflating factors that might be caused by differences in age of the pedigrees themselves confounding the analysis that possibly could explain genetic driven differences rather than age driven differences. One sided p-values were computed for each component. For direct comparison with the Kim et al. analysis of the kidney data we used a consistent model that would test the null hypothesis of no effect of chronological age on gene expression measurement.

**Sex association to gene expression levels**

For the analysis of genes that are differentially expressed between males and females we applied a simple regression model which translates to the standard t-test.

(2) $$Y_{ij} = \beta_{0j} + \beta_{1j} Sex_j + \varepsilon$$

One sided p-values were computed since direction of effect is known.

**Gene Set Enrichment Analysis of differentially expressed sex genes**

We used GSEA implemented in GSEA-P –JAVA version of the software Gene Set Enrichment Analysis [33]. Our goal was to understand what gene sets were enriched for higher expression in males and higher expression in females. Thus, we used the weighted version with Illumina mapping and all four populations.

We also applied Fisher's method to combine the four separate analyses.

Fisher's method is a meta-analysis technique for combining extreme value probabilities from each test into one test statistic having a chi-square distribution: Suppose there are n independent tests with independent continuous test statistics and suppose the null hypotheses are all true, so that the p-values $p_j$ are i.i.d. U[0.1]. Then $\ln(p_j)$ are i.i.d. standard exponential variables and 2ln($p_j$) are i.i.d. exponential with parameter ½, i.e. chi-squared with 2 d.f. Thus $-2\sum_{j=1}^{n}\ln(p_j)$ is chi-squared with 2n degrees of freedom. Using that we can decide when to reject the null hypothesis that all the separate hypotheses are true.

**Association Analyses**

The association analysis was performed after applying quantile normalization to the set of 23,880 gene expression measurements reported [4] with the 69 individuals with genetic and gene expression overlap. SNP genotypes were obtained from build 21 HapMap Phase II. Genomic positions of SNPs were translated to build 36 of the human genome. Association analysis was restricted to SNPs with minor allele frequency above 15%, hence resulting in 1,564,089 SNPs.

**Cis-associations**

For a focus on cis-acting variants we concentrated on a 5MB (+/- 2.5 MB) window of the start and end sites of the 60mer probe sequences. We selected 100 random SNP variants across the genome with minor allele frequency (MAF) > 0.15 for genomic control variance inflation factor estimate calculations under the assumption that 100 randomly selected loci across the genome would be representative of 100 null loci.

**Correcting family bias introduced in the genetics of gene expression analysis**

We use the variance inflation factor robust estimator outlined by Roeder et al. [38,39,40] to control for overdispersion in our test statistic distribution for all of our genetic models. Let Y be the quantitative trait that is influenced by the genotypes at numerous loci. To test if a single locus is associated with the phenotype we work with the model

(3)
$$Y_{ij} = \beta_{0j} + \beta_{1j} Genotype_j + \varepsilon$$

We test the null hypothesis whether the slope $b_1$ is 0. We can define and let $r = \text{Cov}[Y_i, Y_k]$ denote the covariance of phenotypes of individuals in the same subpopulation as outlined in [38]. The usual estimator of the parameter of interest is the least square estimate and among the factors that perturb the distribution of the least square estimate from that expected in typical regression setting include (i) positive correlation among subjects where the variance is increased over that expected under the independence model. When bias is also introduced it results in additional over dispersion [39] which is automatically corrected for by Genomic Control [38]. If we define the standard error (*SE*) term that would be obtained by assuming that the phenotype observations are independent as indicated in [39] we can estimate $b_1$ at *M* loci. A subset of loci that can be designated as null loci, or any set of loci for which the majority are not associated with the quantitative trait suffice to proceed with the genomic control procedure. We can compute our *T* statistic in which family structure or any inherent bias is ignored by estimated b/*SE*. Under the null hypothesis and for large sample sizes *T* is approximately distributed $N(0,l)$, where $l = h^2 + t^2$, in which $h^2$ is proportional to the square of the expected bias of the test statistic and $t^2$ is the increase in variance due to the correlation among subjects. Then, $T_k^2 / l$ is distributed as $c_1^2$. The inflation factor l can be robustly estimated using

(4)
$$\hat{\lambda} = \{median(T_1^2, \ldots, T_M^2)/.456\}.$$

Then we compare $T_k^2$ / estimated(l) with $c_1^2$ to determine whether the locus is significantly associated with the quantitative trait.

**Genomic control for full model:**

To test whether or not adding the age and sex terms increased power to detect cis associations we proceeded with the method proposed in Roeder et al. [39]. The linear regression model

(5) $$Y_{ij} = \beta_{0j} + \beta_{1j}Genotype_j + \beta_{2j}Age_j + \beta_{3j}Sex_j + \varepsilon$$

where we are interested in the estimate of $b_1$. Assuming independent observations SE[$b_1$] is the standard error of the beta estimate computed when fitting the model, $t^2$ is the inflation factor obtained when fitting with a simple additive model as in equation (3), and $H$ is a small positive term that accounts for a slight decrease in the inflation term obtained when fitting equation (5) rather than equation (3). Devlin et al. demonstrate that $t^2 - H$ remains constant across the genome in simulations and therefore Genomic Control applies when fitting a more complex model, and inflation factor is now $t^2 - H$ [39]. To test for multiple effects simultaneously an F-test is appropriated based upon sum of squares where

(6) $$F = (SSE(reduced) - SSE(full))/(df \times MSE(full))$$,

where $df$ is the difference in degrees of freedom between the full and reduced models. Since we are interested in a single locus with multiple terms in the model and we want to asses $H$ in the genome then $df = 1$. Consequently, by simply computing the F-statistic in (6) for the loci under investigation as well as a set of null loci estimating l is based upon the median of the tests obtained from the null loci and adjusting the F-statistic for the variance inflation is done by dividing by l. Since $df = 1$ in this case and we are not testing multiple loci at one $df$. Our lambda estimate is

(7) $$\hat{\lambda} = \{median(F_1, ..., F_M)/.456\}.$$

Then, we evaluate the genotype effect as such.

**Age by genotype interaction model**

Genome-wide age by genotype interaction model for APOE expression level was done via:

(4) $$Y_{ij} = \beta_{0j} + \beta_{1j}Age_j + \beta_{2j}Genotype_j + \beta_{3j}Age_j \times Genotype_j + \varepsilon$$

A sex by genotype term was excluded since empirical data [ref table with association corrections] suggests that the intersection between age-associated and sex-associated transcripts is very small and added degrees of freedom in a model would be overly conservative.

**Transcript Genomic Locations**

We used all 23,880 probes in this analysis. Absolute genomics positions of transcripts were obtained by querying the 60mer sequences against the reference human genome nucleotide sequence NCBI build 36 using the MegaBLAST search program [37] SNP positions were mapped to DB SNP build 126 and its complementary human genome nucleotide reference build 36. These positions were used to determine the absolute distances between the SNP and transcript.

**Acknowledgements**

We thank Roman Yelensky and Dan Arlow for helpful comments.

**Author contributions.** MR was supported by an Eloranta Fellowship from MIT and NIH grant U54 CA121852.

**Gene Expression Datasets:**

**Omnibus Accession IDs:** GSE1726

**GeneVar:** http://www.sanger.ac.uk/humgen/genevar/

**SMD Kidney Expression Data:**

**http://smd.stanford.edu/cgi-bin/publication/viewPublication.pl?pub_no=409**

**Electronic Resources:**

**PLINK:** http://pngu.mgh.harvard.edu/~purcell/plink/

**Haploview:** http://www.broad.mit.edu/mp/haploview/

**GeneVar:** http://www.sanger.ac.uk/humgen/genevar/

**Supplementary Information:**

**Companion Website**

http://www1.cs.columbia.edu/~itsik/projects/FactorsAffectingGeneExpression/rivasdalypeer.html

**Figure Legends**

**Figure 1**. (a) Age distribution of CEPH indivduals analyzed in this study. (b) P-P plot for association with age across all genes reveals large excess of age-associated genes in EBV transformed cell line with extended pedigree expression data, (c) Overlap of age associated genes between kidney and EBV cell lines: The sets of genes whose expression level increases with age significantly overlap across both datasets. (d) Clustering expression signatures of genes across the available set of individuals revealse a tight cluster of age-associated genes including APOE.

**Figure 2.** Frequency of p-value for association with sex across different HapMap populations. In CEU (a), the distribution of p-values is consistent with the null expectation In YRI (b), JPT (c), and CHB (d) there is a significant excess of sex-associated genes.

**Figure 3.** Genomewide p-values for association of HapMap SNPs to APOE expression. (a) P-values for significance of the regression term for Age by genotype interaction contributing to APOE expression, (b) P-values for an additive model for association to APOE expression.

**Supplementary Figure 1.** Pie chart of cis variants discovered under different models: without adjusting for Age and Sex, fitting $Y_{ij} = B_{oj}+B_{1j}\times\text{Gentoype}+\varepsilon$ and with adjustment for age and sex as covariates, fitting $Y_{ij} = B_{oj}+B_{1j}\times\text{Gentoype}+B_{2j}\times\text{Age}+B_{3j}\times\text{Sex}+\varepsilon$. 339 cis acting variants are discovered by applying either model and then applying genomic control. 138 additional cis acting variants are discovered by adjusting for age and sex and another set of additional 155 cis acting variants are accepted by adjusting for age and sex.

Page 27 of **27**